\documentclass{article}
\usepackage{spconf,amsmath,graphicx}
\usepackage[utf8]{inputenc} % Swedish letters
\usepackage{float}

\title{Generating fMRI volumes from T1-weighted volumes using 3D CycleGAN}
\name{David Abramian$^{13}$  \qquad Anders Eklund$^{123}$}

\address{$^{1}$ Division of Medical Informatics, Department of Biomedical Engineering\\
    $^{2}$ Division of Statistics and Machine learning, Department of Computer and Information Science\\
    $^{3}$ Center for Medical Image Science and Visualization\\
    Linköping University, Linköping, Sweden}
\begin{document}
%\ninept
%
\maketitle

\begin{abstract}
Registration between an fMRI volume and a T1-weighted volume is challenging, since fMRI volumes contain geometric distortions. Here we present preliminary results showing that 3D CycleGAN can be used to synthesize fMRI volumes from T1-weighted volumes, and vice versa, which can facilitate registration.  
\end{abstract}

\vspace{-0.5cm}
\section{Introduction}
\label{sec:intro}

Preprocessing of fMRI data is a difficult task, due to head motion and geometric distortions. For example, to perform registration between anatomical (T1) space and fMRI space is difficult due to the contrast difference. The fact that fMRI volumes contain geometric distortions further complicates the registration, since these distortions are normally not present in the anatomical volumes. It is also difficult to evaluate how good a T1-fMRI registration is, mainly due to the distortions, and thereby it is also difficult to compare different fMRI softwares. If it would be possible to synthesize an fMRI volume from a T1-weighted volume, the two volumes would be perfectly registered and can thereby be used as ground truth for evaluating registration algorithms. If the synthetic fMRI image does not contain distortions, it can be used for distortion correction through non-linear registration~\cite{gu2019}. 

Generative adversarial networks (GANs)~\cite{goodfellow2014generative} can today produce very realistic synthetic images~\cite{karras2017progressive}. GANs use adversarial training, where a generator creates an image from noise and a discriminator classifies each image as synthetic or real. During the training, the generator becomes better at generating realistic images, and the discriminator becomes better at discriminating images as synthetic or real. GANs can broadly be divided into noise-to-image GANs~\cite{goodfellow2014generative,karras2017progressive}, which produce an image from a noise vector, and image-to-image GANs, which produce an image from another image (image to image translation)~\cite{isola2017image,zhu2017unpaired}. In medical imaging, GANs have already been used for a number of applications, such as segmentation, synthesis, registration, reconstruction and detection~\cite{yi2019generative}. For image to image translation, CycleGAN~\cite{zhu2017unpaired} is one of the most popular choices. It has previously been used to generate T2-weighted images from T1-weighted images~\cite{welander2018generative}, and to generate diffusion MRI scalar maps from T1-weighted images~\cite{gu2019}, but we are not aware of any application of CycleGAN for T1 - fMRI translation.

In this work, we evaluate if CycleGAN can be used to synthesize realistic fMRI volumes from T1-weighted volumes. Compared to our previous work~\cite{welander2018generative,gu2019,abramian2019refacing} which used a 2D CycleGAN, we here use a 3D CycleGAN to take advantage of the 3D correlations present in the data.

\vspace{-0.25cm}
\section{Theory}
\label{sec:format}

\vspace{-0.2cm}
\subsection{CycleGAN}

CycleGAN~\cite{zhu2017unpaired} can be trained using two unpaired groups of images or volumes, to translate data between domain A and domain B (e.g. T1 and fMRI). CycleGAN consists of four main components, two generators ($G_{A2B}$, $G_{B2A}$) and two discriminators ($D_A$ and $D_B$). The two generators generate domain A/B images based on domain B/A images. The two discriminators discriminate each image as synthetic or real. The unsupervised training is regularized by the cycle consistency
\begin{align}
  G_{B2A}(G_{A2B}(I_A)) &\approx I_A, \\
  G_{A2B}(G_{B2A}(I_B)) &\approx I_B,
\end{align}
where $I_A$ and $I_B$ are two images from domain A and B. 

\vspace{-0.15cm}
\subsection{2D vs 3D}

CycleGAN was initially proposed for 2D images, but conditional GANs have also been applied in 3D. Nie et al.~\cite{nie2017medical} used a 3D conditional GAN to synthesize CT from MR, and worked on subvolumes of 32 x 32 x 32 voxels. Näppi et al.~\cite{nappi2019cycle} used 3D CycleGAN for virtual bowel cleansing in CT colonography, and worked on volumes of 96 x 96 x 96 voxels. Yu et al.~\cite{yu20183d} used a 3D conditional GAN to synthesize FLAIR images from T1-weighted images, to improve brain tumour segmentation. The conditional GAN processed volumes of size 240 x 240 x 155 voxels using subvolumes of 128 x 128 x 128 voxels. 

For 3D volumes there are several different options. The most straight forward approach is to train a 2D CycleGAN and then apply it to every slice in a volume. Such an approach will not take advantage of the available 3D information, and can lead to discontinuities between the slices. Training a 3D CycleGAN can be more computationally demanding, since a (non-separable) 3D convolution involves more calculations compared to a 2D convolution. Another challenge of a 3D CycleGAN is the memory consumption, since training a CycleGAN involves training four convolutional neural networks (CNNs) at the same time (two generators and two discriminators), and many filter responses need to be stored for every layer of each CNN. 

Training in 3D can be performed with full size volumes (e.g. 128 x 128 x 128 or 256 x 256 x 256 voxels), or with subvolumes (e.g. 32 x 32 x 32 voxels). Using full size volumes means that only a single training example is available for each subject, and to fit several hundred filter responses in GPU memory can be difficult for large volumes. Using subvolumes can lead to a large number of training examples for each subject (it is possible to extract a 32 x 32 x 32 subvolume from a 128 x 128 x 128 volume in many different ways), and the memory consumption is also substantially reduced. If the network is fully convolutional, the trained network can then be applied to a volume of any size. The drawback of using subvolumes is that the performance may be reduced, since the network does not see the full volume. Since it in our case is possible to use a minibatch of 1 for volumes of 152 x 180 x 120 voxels we focused on the full size approach. 

\vspace{-0.4cm}
\section{Data}
\label{sec:pagestyle}
\vspace{-0.2cm}

We used fMRI and T1-weighted volumes from the 1000 functional connectomes project~\cite{biswal2010toward}, and focused on the Beijing and Cambridge datasets since they consist of data from 198 subjects each. For Beijing, the T1 volumes have a varying size and voxel size of 1.3 x 1.0 x 1.0  mm$^3$, while the fMRI data have a size of 64 x 64 x 33 voxels and a voxel size of 3.125 x 3.125 x 3.6 mm$^3$. For Cambridge, the T1 volumes have a varying size and a voxel size of 1.2 x 1.2 x 1.2 mm$^3$, while the fMRI data have a size of  72 x 72 x 47 voxels and a voxel size of 3 x 3 x 3 mm$^3$. We trained the network with 160 subjects, and used the remaining 38 subjects for testing. 

%We used diffusion and T1 images from the Human Connectome Project (HCP)\footnote{Data collection and sharing for this project was provided by the Human Connectome Project (U01-MH93765) (HCP; Principal Investigators: Bruce Rosen, M.D., Ph.D., Arthur W. Toga, Ph.D., Van J.Weeden, MD). HCP fund- ing was provided by the National Institute of Dental and Craniofacial Re- search (NIDCR), the National Institute of Mental Health (NIMH), and the National Institute of Neurological Disorders and Stroke (NINDS). HCP data are disseminated by the Laboratory of Neuro Imaging at the University of Southern California.} \cite{van2013wu,glasser2013minimal}  for 1065 subjects. The data were collected using a customized Siemens 3T Connectom scanner. The diffusion data were acquired with 3 different b-values (1000, 2000, and 3000 $s/mm^2$) and have already been pre-processed for gradient nonlinearity correction, motion correction and eddy current correction. The diffusion data consist of 18 non-diffusion weighted volumes (b = 0) and 90 volumes for each b-value, which yields 288 volumes of $145 \times 174 \times 145$ voxels with an 1.25 mm isotropic voxel size. The T1 data was acquired with a $0.7 \times 0.7 \times 0.7$ mm isotropic voxel size and then downsampled to the same resolution as the diffusion data.

\vspace{-0.4cm}
\section{Methods}
\vspace{-0.2cm}
\subsection{Preprocessing}

CycleGAN can be trained using unpaired (and unregistered) images, but we registered the fMRI and T1 data to the MNI 152 brain template to facilitate training and evaluation. We first used the epi\_reg script in FSL, which uses boundary based registration~\cite{greve2009accurate} for fMRI to T1 registration. We then used flirt in FSL~\cite{jenkinson2002improved} to linearly register all T1 volumes to the MNI 152 template, and finally combined the transformations to transform the fMRI data to MNI space. To reduce training time and memory consumption, we cropped the MNI 152 1 mm brain template from 182 x 218 x 182 to 152 x 180 x 120 voxels. Processing scripts are available on Github\footnote{https://github.com/wanderine/fMRI\_GAN}~\cite{eklund2017reply}.

\vspace{-0.15cm}
\subsection{Data augmentation}

The number of available training volumes is only 160. To increase the number of training volumes we performed data augmentation by applying 10 random 3D rotations to each of the 160 volumes, to achieve a total of 1760 training volumes. The random rotations were generated from a normal distribution with a mean of 0 and a standard deviation of 10 degrees.

\vspace{-0.25cm}
\subsection{CycleGAN}

We based our 3D CycleGAN on the 2D Keras implementation provided in \cite{welander2018generative}. We replaced all 2D convolutions with 3D convolutions, and reduced the number of residual layers from 9 to 6, as recommended for smaller images~\cite{zhu2017unpaired}. The architecture for the generator is given in Table~\ref{table:generator}, and the architecture for the discriminator is given in Table~\ref{table:discriminator}. The volume size for the patchGAN, which classifies each patch (subvolume) as real or synthetic, was reduced from 70 x 70 x 70 to 51 x 51 x 51, since our volumes are rather small compared to the images in the original CycleGAN paper~\cite{zhu2017unpaired}. Training the 3D CycleGAN for 200 epochs, using a minibatch size of 1, took 24 days on an Nvidia Tesla V100 graphics card with 32 GB of memory. 

%We used two widely used measures from the literature \cite{wolterink2017deep,emami2018generating}, the correlation coefficient (CC) and the structural similarity (SSIM), to quantify the accuracy of the image translation. The CC measures the degree of global correlation between two images and SSIM can measure local structural similarity between two images.
%The CC is defined as \cite{lee1988thirteen}
%\begin{align}
 % CC=\frac{\sum_{n}^{}\left [I_A(n)-\bar{I}_A\right ]\left [I_B(n)-\bar{I}_B\right ]}{\sqrt{\sum_{n}^{}\left [I_A(n)-\bar{I}_A\right ]^2 \sum_{n}^{}\left [I_B(n)-\bar{I}_B\right ]^2}},
%\end{align}
%where $I_A$ and $I_B$ are images A and B, $\bar{I}_A$ and $\bar{I}_B$ are the mean of A and B, $n$ is the pixel index.
%The SSIM quantifies the degree of similarity of two images based on the impact of three characteristics: luminance, contrast and structure. The SSIM of pixel $(x,y)$ in images A and B can be calculated as \cite{wang2004image}
%\begin{align}
 % SSIM(x,y) = \frac{(2 \mu_{w_A} \mu_{w_B} +c_1)(2 \sigma_{w_Aw_B}+c_2)}{(\mu_{w_A}^2+\mu_{w_B}^2+c1)(\sigma_{w_A}^2+\sigma_{w_B}^2+c_2)},
%\end{align}
%where $w_A$ and $w_B$ are local neighborhoods centered at $(x,y)$ in images A and B, $\mu_{w_A}$ and $\mu_{w_B}$ are the local means, $\sigma_{w_A}$ and $\sigma_{w_B}$ are the local standard deviations, $\sigma_{w_Aw_B}$ is the covariance, $c_1$ and $c_2$ are two variables to stabilize the division. The mean SSIM ($MSSIM=SSIM/N_{voxel}$) within the brain area can be used as a global measure of the similarity between the synthetic image and the ground truth.

\vspace{-0.5cm}
\section{Results}
\label{sec:typestyle}

\begin{figure*}
\begin{minipage}[b]{1.0\linewidth}
  \centering
  \centerline{\includegraphics[width=1.05\textwidth]{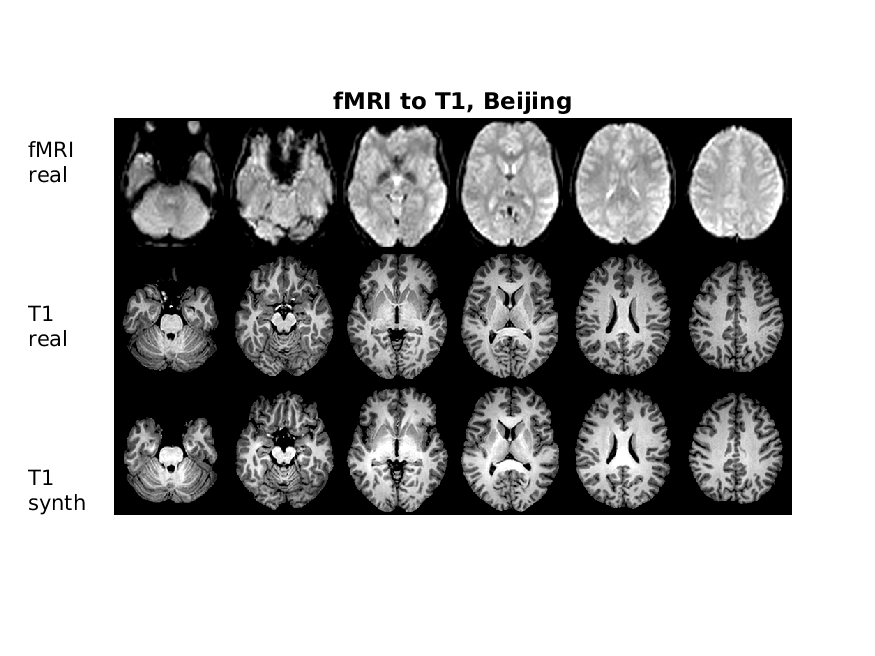}}
\end{minipage}
\caption{fMRI to T1 image translation results for 6 slices of a test subject, for Beijing data. First row: real fMRI images, second row: real T1 images, third row: synthetic T1 images.}
\label{figure1}
\end{figure*}

\begin{figure*}
\begin{minipage}[b]{1.0\linewidth}
  \centering
  \centerline{\includegraphics[width=1.05\textwidth]{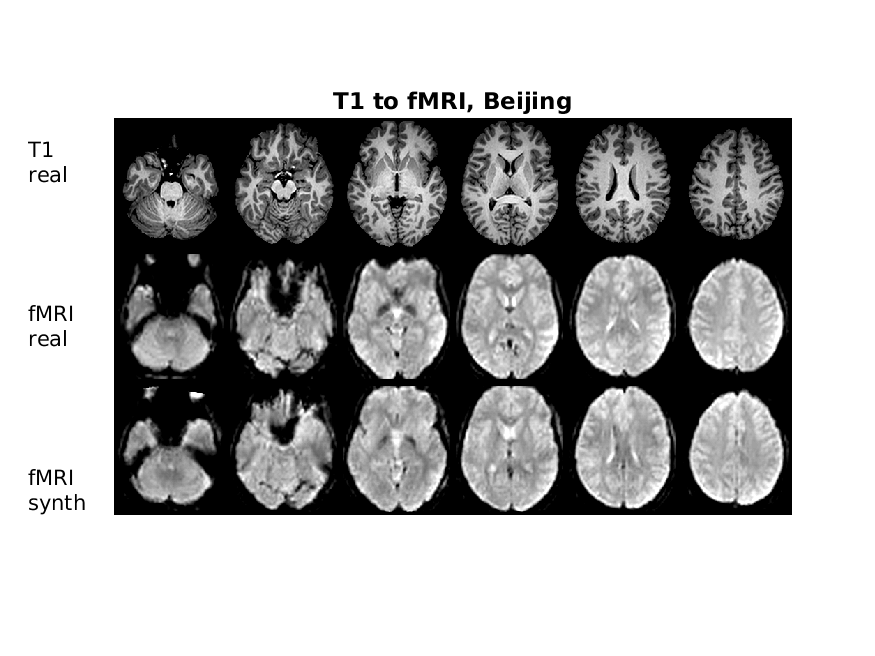}}
\end{minipage}
\caption{T1 to fMRI image translation results for 6 slices of a test subject, for Beijing data. First row: real T1 images, second row: real fMRI images, third row: synthetic fMRI images.}
\label{figure2}
\end{figure*}

\begin{figure*}
\begin{minipage}[b]{1.0\linewidth}
  \centering
  \centerline{\includegraphics[width=1.05\textwidth]{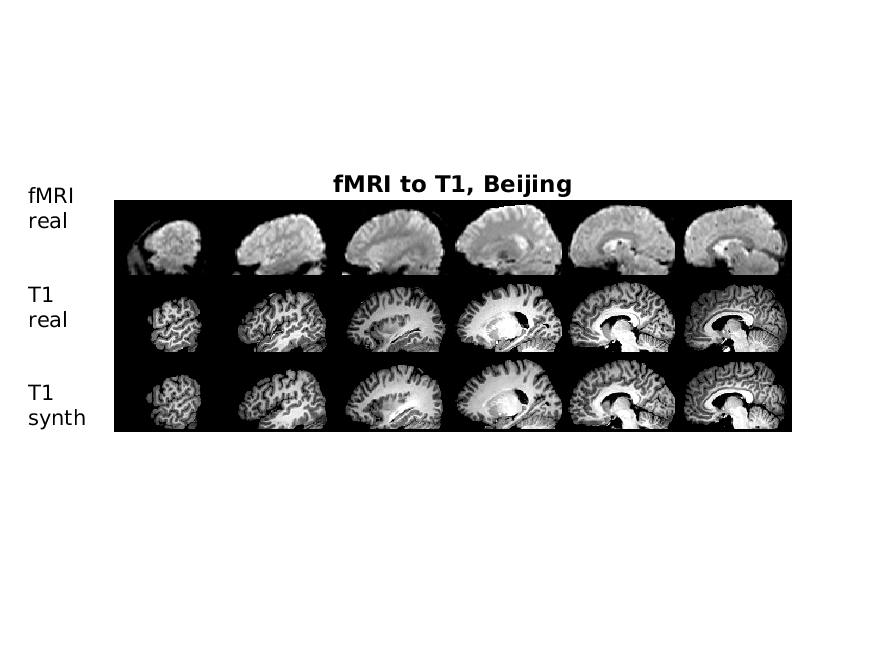}}
\end{minipage}
\caption{fMRI to T1 image translation results for 6 slices of a test subject, for Beijing data. First row: real fMRI images, second row: real T1 images, third row: synthetic T1 images.}
\label{figure3}
\end{figure*}

\begin{figure*}
\begin{minipage}[b]{1.0\linewidth}
  \centering
  \centerline{\includegraphics[width=1.05\textwidth]{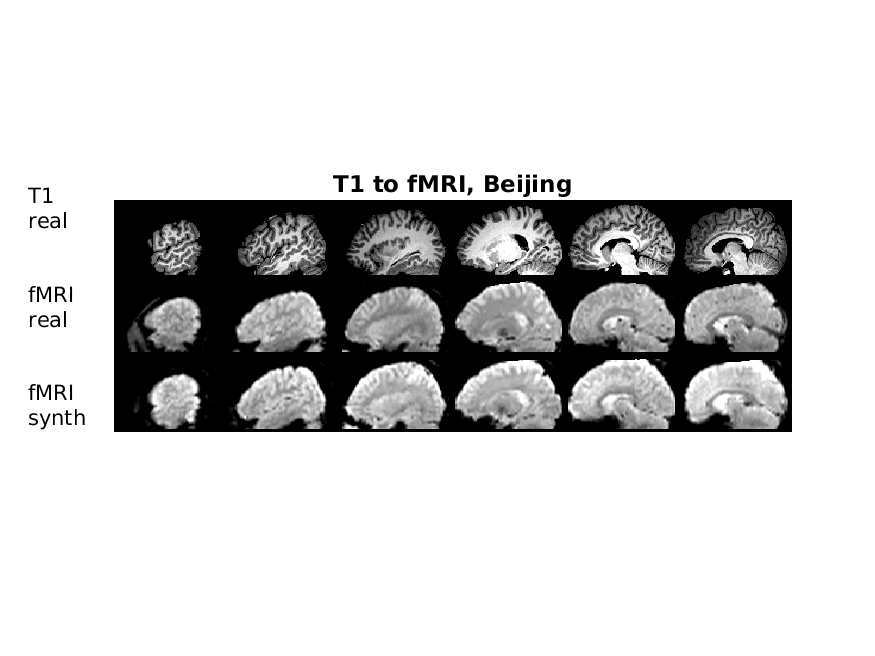}}
\end{minipage}
\caption{T1 to fMRI image translation results for 6 slices of a test subject, for Beijing data. First row: real T1 images, second row: real fMRI images, third row: synthetic fMRI images.}
\label{figure4}
\end{figure*}

\begin{figure*}
\begin{minipage}[b]{1.0\linewidth}
  \centering
  \centerline{\includegraphics[width=1.05\textwidth]{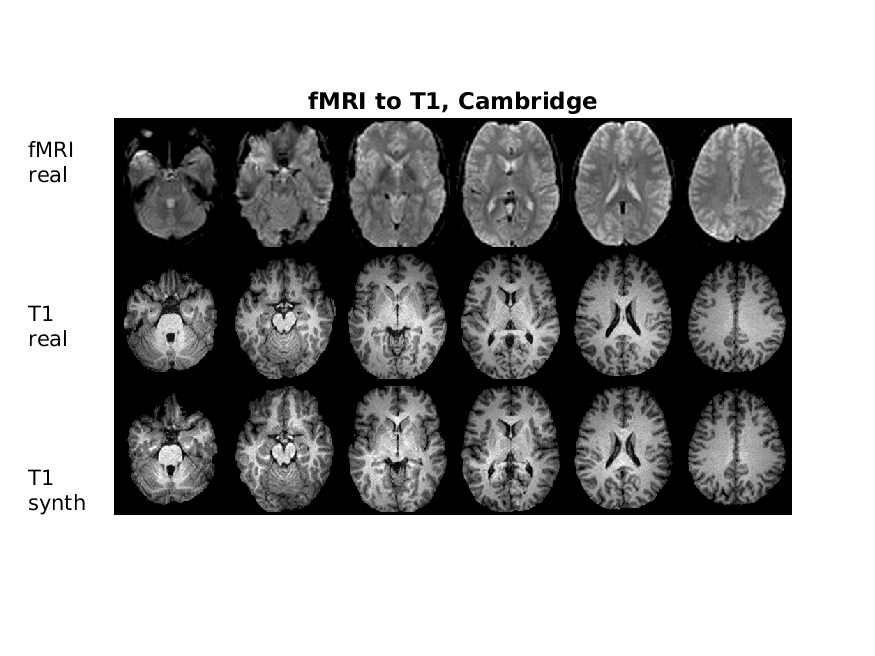}}
\end{minipage}
\caption{fMRI to T1 image translation results for 6 slices of a test subject, for Cambridge data. First row: real fMRI images, second row: real T1 images, third row: synthetic T1 images.}
\label{figure5}
\end{figure*}

\begin{figure*}
\begin{minipage}[b]{1.0\linewidth}
  \centering
  \centerline{\includegraphics[width=1.05\textwidth]{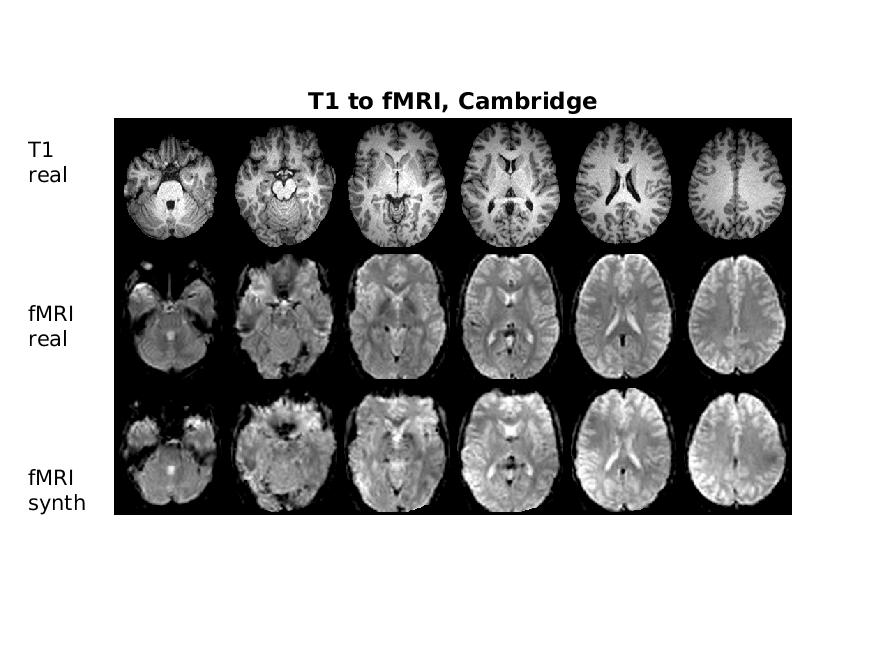}}
\end{minipage}
\caption{T1 to fMRI image translation results for 6 slices of a test subject, for Cambridge data. First row: real T1 images, second row: real fMRI images, third row: synthetic fMRI images.}
\label{figure6}
\end{figure*}

\begin{figure*}
\begin{minipage}[b]{1.0\linewidth}
  \centering
  \centerline{\includegraphics[width=1.05\textwidth]{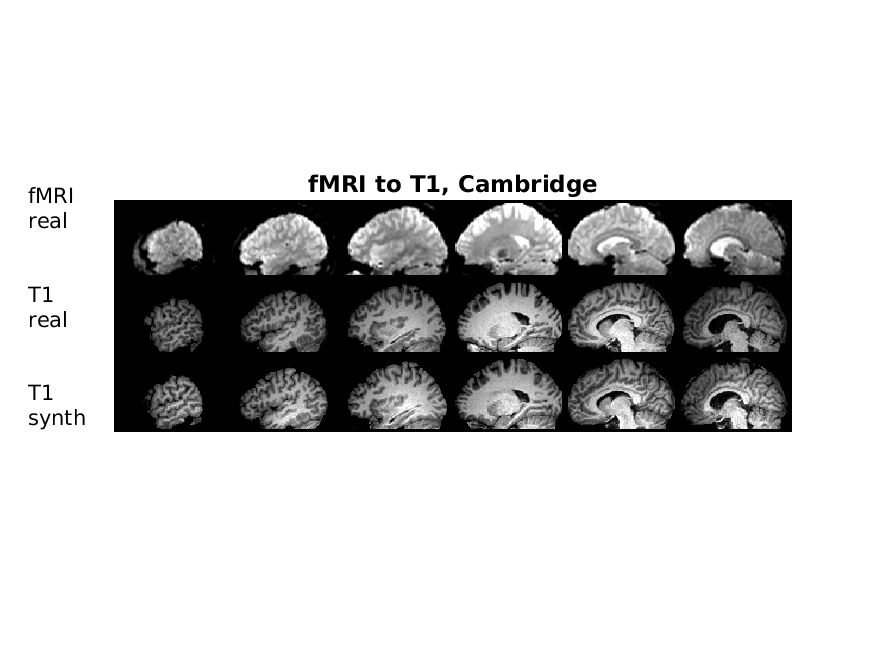}}
\end{minipage}
\caption{fMRI to T1 image translation results for 6 slices of a test subject, for Cambridge data. First row: real fMRI images, second row: real T1 images, third row: synthetic T1 images.}
\label{figure7}
\end{figure*}

\begin{figure*}
\begin{minipage}[b]{1.0\linewidth}
  \centering
  \centerline{\includegraphics[width=1.05\textwidth]{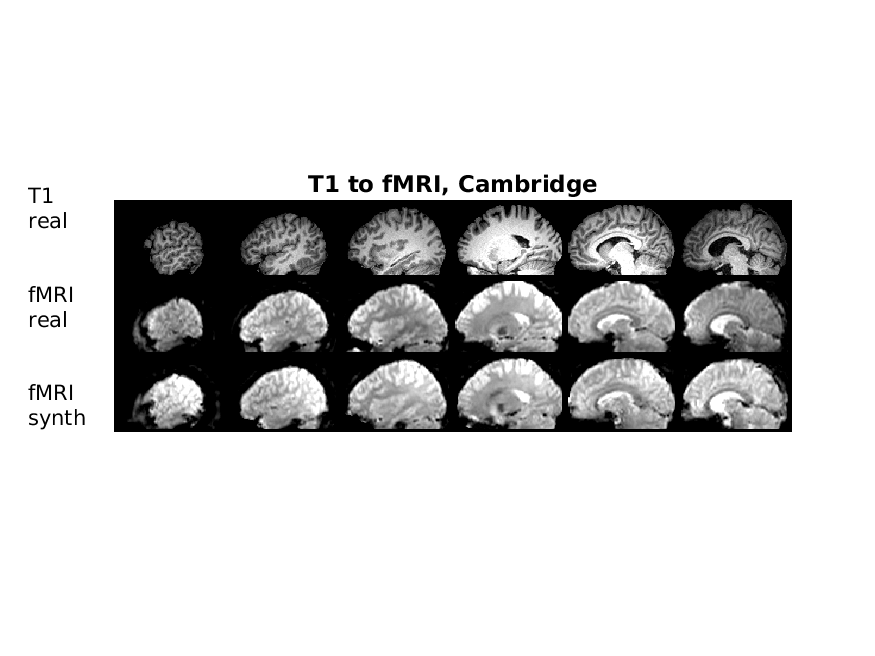}}
\end{minipage}
\caption{T1 to fMRI image translation results for 6 slices of a test subject, for Cambridge data. First row: real T1 images, second row: real fMRI images, third row: synthetic fMRI images.}
\label{figure8}
\end{figure*}

Figures~\ref{figure1} - ~\ref{figure4} show results for Beijing data, and Figures~\ref{figure5} - ~\ref{figure8} show results for Cambridge data. For slices in the upper part of the brain the fMRI-T1 transformation is not so difficult (mainly a change in contrast), while a rather non-liner transformation must be learned for slices in the lower part of the brain (due to severe distortions). 

\vspace{-0.3cm}
\section{Conclusion}
\vspace{-0.1cm}

We have demonstrated that 3D CycleGAN can be used to synthesize fMRI images from T1-weighted images, and vice versa. We have here focused on transforming a full volume, rather than working on subvolumes, but in future work we will make a comparison between the two solutions. 

In future work we will also investigate the performance of T1-fMRI registration for different fMRI softwares, by generating a synthetic fMRI volume from the T1-weighted volume, and then apply different transformations that should be found by the registration algorithms.

\vspace{-0.25cm}
\section*{Acknowledgements}

This study was supported by Swedish research council grant 2017-04889. Funding was also provided by the Center for Industrial Information Technology (CENIIT) at Linköping University, Analytic Imaging Diagnostics Arena (AIDA) and the ITEA3 / VINNOVA funded project ”Intelligence based iMprovement of Personalized treatment And Clinical workflow supporT” (IMPACT).

\begin{table*}[htb]
\scriptsize
\caption{Architecture used for the generator in our 3D CycleGAN.}
\begin{center}
\begin{tabular}{|c|c|c|c|c|c|}
\hline  
\textbf{\normalsize Layer}  & \textbf{\normalsize Layer type}  & \textbf{\normalsize Number of filters} & \textbf{\normalsize Filter size} & \textbf{\normalsize Stride} & \textbf{\normalsize Activation function}    \\[0.2ex]
\hline 
\normalsize 1 & \normalsize 3D Convolution & \normalsize 32 & \normalsize 7 & \normalsize 1 & \normalsize ReLU \\

\normalsize 2 & \normalsize Downsampling (3D Convolution) & \normalsize 64 & \normalsize 3 & \normalsize 2 & \normalsize ReLU \\

\normalsize 3 & \normalsize Downsampling (3D Convolution) & \normalsize 128 & \normalsize 3 & \normalsize 2 & \normalsize ReLU \\

\normalsize 4 - 9 & \normalsize Residual (3D Convolution) & \normalsize 128 & \normalsize 3 & \normalsize 1 & \normalsize None \\

\normalsize 10 & \normalsize Upsampling (3D Transpose Convolution) & \normalsize 64 & \normalsize 3 & \normalsize 2 & \normalsize ReLU \\

\normalsize 11 & \normalsize Upsampling (3D Transpose Convolution) & \normalsize 32 & \normalsize 3 & \normalsize 2 & \normalsize ReLU \\

\normalsize 12 & \normalsize 3D Convolution & \normalsize 1 & \normalsize 7 & \normalsize 1 & \normalsize tanh \\

\hline
\end{tabular}
\end{center}
\label{table:generator}
\end{table*}

\begin{table*}[htb]
\scriptsize
\caption{Architecture used for the discriminator in our 3D CycleGAN.}
\begin{center}
\begin{tabular}{|c|c|c|c|c|c|}
\hline
\textbf{\normalsize Layer}  & \textbf{\normalsize Layer type}  & \textbf{\normalsize Number of filters} & \textbf{\normalsize Filter size} & \textbf{\normalsize Stride} & \textbf{\normalsize Activation function}    \\[0.2ex]
\hline
\normalsize 1 & \normalsize 3D Convolution & \normalsize 64 & \normalsize 4 & \normalsize 2 & \normalsize LeakyReLU (0.2) \\

\normalsize 2 & \normalsize 3D Convolution & \normalsize 128 & \normalsize 4 & \normalsize 2 & \normalsize LeakyReLU (0.2) \\

\normalsize 3 & \normalsize 3D Convolution & \normalsize 256 & \normalsize 4 & \normalsize 1 & \normalsize LeakyReLU (0.2) \\

\normalsize 4 & \normalsize 3D Convolution & \normalsize 512 & \normalsize 4 & \normalsize 1 & \normalsize LeakyReLU (0.2) \\

\normalsize 5 & \normalsize 3D Convolution (PatchGAN) & \normalsize 1 & \normalsize 4 & \normalsize 1 & \normalsize Sigmoid \\

\hline
\end{tabular}
\end{center}
\label{table:discriminator}
\end{table*}

% To start a new column (but not a new page) and help balance the last-page
% column length use \vfill\pagebreak.
% -------------------------------------------------------------------------
%\vfill
%\pagebreak
% \clearpage

% References should be produced using the bibtex program from suitable
% BiBTeX files (here: strings, refs, manuals). The IEEEbib.bst bibliography
% style file from IEEE produces unsorted bibliography list.
% -------------------------------------------------------------------------
\clearpage
\bibliographystyle{IEEEbib}
\bibliography{refs}

\end{document}